\documentclass[12pt, epsfig, preprint2]{aastex}

\shorttitle{H$\alpha$ Filament in SN1006} \shortauthors{Raymond et al.}

\begin{document}

\title{The pre-shock gas of SN1006 from HST/ACS observations}
\author{J.C. Raymond\altaffilmark{1}, K.E. Korreck\altaffilmark{1}, Q.C. Sedlacek\altaffilmark{1},
W. P. Blair \altaffilmark{2}, P. Ghavamian\altaffilmark{2}, R. Sankrit\altaffilmark{3}}

\altaffiltext{1}{Harvard-Smithsonian Center for
Astrophysics, 60 Garden Street, Cambridge, MA 02138}
\altaffiltext{2}{The Johns Hopkins University, Baltimore, MD}
\altaffiltext{3}{Space Sciences Laboratory, University of California, Berkeley, CA}

\begin{abstract}

We derive the pre-shock density and scale length along the line 
of sight for the collisionless shock from a deep HST image that 
resolves the H$\alpha$ filament in SN1006 and updated model 
calculations.   The very deep ACS high-resolution image of the 
Balmer line filament in the northwest (NW) quadrant shows that 
0.25 $\le~n_0~\le$ 0.4 $\rm cm^{-3}$ and that the scale along 
the line of sight is about $2 \times 10^{18}~\rm cm$, while 
bright features within the filament correspond to ripples with 
radii of curvature less than 1/10 that size.  The derived 
densities are within the broad range of earlier density 
estimates, and they agree well with the ionization time scale 
derived from the Chandra X-ray spectrum of a region just behind
the optical filament.  This provides a test for widely used 
models of the X-ray emission from SNR shocks.  The scale and 
amplitude of the ripples are consistent with expectations for 
a shock propagating though interstellar gas with $\sim$ 20\% 
density fluctuations on parsec scales as expected from studies 
of interstellar turbulence.  One bulge in the filament 
corresponds to a knot of ejecta overtaking the blast wave, 
however.  The interaction results from the rapid deceleration of 
the blast wave as it encounters an interstellar cloud.

\end{abstract}

\keywords{ISM:individual(SN1006)--supernova remnants--shock waves--optical:ISM}

\section{Introduction}

SN1006 (G327.6+14.6) is one of the best SNRs for studying the physics of collisionless
astrophysical shocks, in particular the acceleration of non-thermal particles.
It is a nearby Type $\textrm{I}$a supernova remnant at a distance of
2.1 kpc (the distance we assume throughout) with a diameter of $\sim$ 18 pc \citep{wink03}
and a shock speed  in the 2500 - 2900 $\rm km~s^{-1}$ range \citep{gha02, hm}. 
The remnant has a high Galactic latitude and modest foreground
reddening, E(B-V)=0.11 $\pm$ 0.02 \citep{sch80}.  

SN1006 has been observed at radio \citep{rey93, moffett},  optical
\citep{gha02,kwc87,smi91}, ultraviolet \citep{ray95, korreck} and X-ray
\citep{koyama, wink03,lon03,bam03,dye04} wavelengths.  Pure Balmer
line filaments were found in the optical by \citet{vdb}. In the radio and X-ray,
the remnant has a limb-brightened shell structure with cylindrical
symmetry around a SE to NW axis probably aligned with the ambient
galactic magnetic field \citep{rey96,jp89}.  The NE shock front of
SN1006 shows strong non-thermal X-ray emission \citep{koyama, dye04}
while the NW shock shows very little non-thermal emission
at radio or X-ray wavelengths.  On the other hand, the pre-shock
density is several times higher in the NW than the NE \citep{korreck}.
Knots of X-ray emission from shocked SN ejecta
are scattered through the interior of the remnant \citep{lon03, vink03}.

A Balmer line filament defines the blast wave in the northwest quadrant of 
SN1006, and the H$\alpha$ profile provides diagnostics for the
shock speed and ion-electron thermal equilibration.  
The H$\alpha$ emission from a Balmer-dominated shock has a two component profile \citep{cr78}. The
broad component is due to charge exchange between neutrals and protons, which
produces a population of neutrals at nearly the post-shock proton temperature.
The narrow component is produced
when cold ambient neutrals pass through the shock and emit line
radiation before charge transfer or ionization occurs.  The ratio
of the broad to narrow flux is sensitive to the electron and ion temperatures.
The FWHM of H$\alpha$ broad component is 2290 $\pm$ 80 km s$^{-1}$, and
models imply a shock speed of v$_{\rm{shock}}= 2890
\pm 100$ km s$^{\rm{-1}}$ for a shock with little
electron-ion equilibration \citep{gha02}.  However, new models incorporating 
some additional physics obtain a lower shock speed of 2509$\pm$111 $\rm km~s^{-1}$
\citep{hm}. 

The pre-shock density, $n_0$, is an important parameter for understanding
the evolution of SN1006, as well as for interpreting the X-ray spectra
in terms of ionization time scale and determining the relative contributions
of shocked ISM and SN ejecta to the X-ray emission.  The density could
also be important for attempts to understand the high ratio of
non-thermal to thermal X-ray emission in this SNR.  Estimates of
$n_0$ cover a wide range, from 0.05-0.1 $\rm cm^{-3}$ based on the
global X-ray emission \citep{hss86} to 1 $\rm cm^{-3}$ based on
interpreting the scale over which the X-rays brighten as the length
scale for ionization of the shocked plasma \citep{win97}.  A related
parameter is the length scale along the line of sight.  The ripples
in the SNR blast wave could in principle result from density
inhomogeneities in the ambient medium or from knots of SN ejecta
overtaking the blast wave.  ISM density fluctuations have been inferred for
a section of the non-radiative shock in the Cygnus Loop \citep{ray03},
while an ejecta knot is clearly the cause of one bulge in the
H$\alpha$ filament of SN1006 \citep{lon03, vink03}.  The issue is
important for estimating the amplitude of interstellar turbulence
on sub-parsec scales and for the interpretation of the distance
between the blast wave and the reverse shock in terms of particle
acceleration \citep{warren}. 

In this paper we use an H$\alpha$ image obtained with the ACS imager on the
Hubble Space Telescope to determine the pre-shock density and the length scale
along the line of sight.  This is possible because the ACS images resolve the
thickness of the narrow zone behind the shock where hydrogen atoms are excited
and ionized, and that thickness scales inversely as the pre-shock density.  We
have computed new models of the H$\alpha$ emissivity as a function of distance
behind the shock taking into account recent results by \citet{hm}.  We
discuss the observations in the next section, then compare with models to derive the
shock parameters.  Section 4 provides a limit on the brightness of any 
shock precursor, compares the densities with other estimates and
discusses the implications for other analyses of SN1006 observations. 
Section 5 summarizes the conclusions.

\section{Observations}

The Hubble Space Telescope's ACS Wide Field Camera (WFC) imaged a full 
field of 202 x 202 arcsec$^2$ at coordinates $\alpha_{2000}$ 
=$15^{\rm{h}}$ $2^{\rm{m}}$ $19.02^{\rm{s}}$, $\delta_{2000}$ =-41$^o$ 
44\arcmin\ 48.4\arcsec\ on 9 orbits from 15-17 February 2006.   Exposures 
with durations of 2,746.0 s, 2,848.0 s, and 2,828.0 s were obtained for 
each subset of 3 orbits, for a combined 25,266 second exposure.  The exposures 
were taken with the F658N H-alpha filter, which has a flat response at wavelengths
above about 6558\AA , drops to half the peak transmission at about 6548 \AA \/ and
20\% of the peak transmission at 6540 \AA .  This means that it passes the narrow component,
all the red wing of the broad component, and about half the blue wing of
the broad component, or about 90\% of the H$\alpha$ emission. 

The image was centered on the position where Ghavamian et al. (2002) obtained
a low dispersion spectrum and Sollerman et al. (2003) obtained a high dispersion
spectrum, so we are able to use the shock speed, neutral fraction
and electron-ion equilibration derived from those observations. 
The position was also observed in the ultraviolet with the
Hopkins Ultraviolet Telescope \citep{ray95, lam96}  and
with FUSE \citep{korreck}, so that we can use the information
derived in those papers about ion-ion thermal equilibration to
constrain model parameters.  The
H$\alpha$ filament also defines the outer boundary of position NW-1
in the X-ray spectral analysis of Long et al. (2003), who 
found the spectrum to be consistent with thermal emission from
shocked interstellar gas.

The raw data were combined and reduced with the standard ACS 
calibration pipeline (CALACS), involving bias/dark-current 
subtraction, flat-fielding, image combination, and cosmic-ray 
rejection (Sirianni et. al. 2005), then 'drizzled' (Fruchter 
and Hook 2002) onto a 0.03 arcsec pixel scale with the task
'multidrizzle' to correct for 
geometric distortion and improve the sampling of the point 
spread function \citep{koekemoer}.

Figure~\ref{xrayhalpha} shows the HST image overlaid on the Chandra X-ray image.
The morphology of the filament is clearly that of a rippled 
sheet seen edge-on, with the bright rims corresponding to tangencies
to the line of sight \citep{hester}.  For the present purposes we are interested
in the simplest tangencies, since those are amenable to modeling.
The bulge near the SW corner of the image is morphologically similar to a larger,
brighter region farther to the SW where a clump of ejecta is overtaking the shock 
\citep{lon03, vink03}, but it shows only a slight X-ray enhancement.  
The shock morphology in the bulge is more complex, so this paper 
concentrates on the smoother regions of the filament.

IDL was used to extract and plot the curve of the shock 
front spanning the drizzled image and to find the approximate direction
perpendicular to the shock.  At each of 40 positions we extracted
the spatial profiles across the shock for a range of angles near
the initial estimate and selected the the profile showing the
narrowest H$\alpha$ peak as the one closest to the shock normal direction.  Many
of the profiles suffer from low signal to noise or from complexity
due to several tangencies to the line of sight.  We have selected
8 profiles with bright, simple H$\alpha$ peaks for further 
analysis.   In particular, we model sections of the trailing
edge of the filament in regions corresponding to sections F, G and H
of the \citet{wink03} proper motion analysis.  Figure~\ref{halpha}
shows the boxes used to extract spatial profiles, starting with
profile 8 in the upper left and ending with profile 29 in the lower
right.  The profiles were extracted using boxes 4.5\arcsec wide and
9\arcsec long.  Their positions and position angles are shown in
Table 1.  The profiles extracted
were similar to those used by \citep{wink03} for
measuring the proper motion of the filament; here, however, the goal was
to determine the width, not the position, of the filament.

\section{Analysis and Results}

To interpret the images we compute models of the H$\alpha$
brightness behind a curved shock, convolve the model intensity distribution
with the ACS point spread function, and compare the models to the observations.
In general, the pre-shock density controls the brightness drop off behind
the peak, because the thickness of the emission region is inversely
proportional to the density.  The radius of curvature of the shock determines the fall off
ahead of the peak for the concave outward filaments that we model here.
The absolute intensity scales approximately as $n_0^2 f_{neut} R^{1/2}$, where
$n_0$ is the pre-shock density, $f_{neut}$ is the pre-shock neutral fraction,
and R is the radius of curvature of the shock.
 
\subsection{Model Calculations}

We start with model calculations similar to those of \citet{lam96}.  Figure \ref{model}
shows the H$\alpha$ emissivity as a function of distance behind the shock for
a 2900 $\rm km~s^{-1}$ shock model with pre-shock density $n_0~=~0.25~\rm cm^{-3}$,
a neutral fraction of 0.1, and a ratio $T_e / T_p~= 0.05$ at the shock front.  The
model follows neutral hydrogen as it passes through the shock and undergoes 
collisional excitation and ionization by protons and electrons as well as charge
transfer with post-shock protons.  The atomic rates are described by \citet{lam96},
and Coulomb collisions slowly transfer energy from the ions to the electrons.  The
radiative transfer involved in the conversion of Ly$\beta$ photons to H$\alpha$
photons, important for the narrow component, is described in \citet{lam96}.  A fraction
of the H$\alpha$ arises from converted Ly$\beta$ photons, and those H$\alpha$
photons are produced over a scale of about 1 Ly$\beta$ mean free path.  That scale
is about 0.1\arcsec, which is small enough compared to the observed filament widths
that we ignore it.

Evidence for a shock precursor has
been reported for a number of non-radiative shocks from low ionization emission
lines \citep{hes94, sol03}, from the velocity widths of their narrow components \citep{smi94, hes94}, 
and from the spatial distribution of the narrow component emission
\citep{lee06}.  The narrow component in SN1006 shows no broadening beyond that
expected for the ambient ISM \citep{sol03}, so we do not include any precursor
emission in the models.

Recently, \citet{hm} investigated the consequences of the sharp decline of the
the charge transfer cross section at speeds above about 2000 $\rm km~s^{-1}$ for the
Balmer line profiles.  They estimated a shock speed of 2509$\pm$111 $\rm km~s^{-1}$
from the Balmer line profile presented by \citet{gha02}, as opposed to the value 
2890$\pm$100 $\rm km~s^{-1}$ found by Ghavamian et al.  Both values apply to the
case of little ion-electron thermal equilibration in the shock.  Strong equilibration
is ruled out by the X-ray spectrum of \citet{lon03} and by the combination of
broad to narrow intensity ratio given by \citet{gha02} and the Ly$\beta$ radiative
transfer calculations of \citet{lam96}.  The smaller shock speed of \citet{hm} would decrease the
distance scale of the H$\alpha$ emission behind the shock.  However, it would also
imply a smaller distance for SN1006 based the combination of the proper motion and the
and shock speed \citep{wink03}, so the angular width of the emission region would be unchanged
to first order.

There is another important implication of the work of \citet{hm} for the present
study.  The rapid drop in charge transfer cross section with increasing
velocity means that neutrals are more likely to undergo charge transfer with protons
moving away from the shock than with protons moving toward the shock.  While the
velocity dependence of the cross section was included in earlier models, the anisotropy
of the resulting H I velocity distribution was not.  Thus the earlier models implicitly
assumed that the broad component neutrals move away from the shock at $V_s$/4, the
same speed as the post-shock protons.  By integrating the product of velocity
times charge transfer cross section, $\sigma_{cx} v$ \citep{redbook},
over Maxwellian distributions at the post shock temperatures, we
find that after 1 charge transfer event the average neutral is moving away from the shock
at 1500 $\rm km~s^{-1}$ behind a 3000 $\rm km~s^{-1}$ shock, rather than the 750 $\rm km~s^{-1}$
of the ionized gas.  For a 2500 $\rm km~s^{-1}$ shock, the neutrals move at 1100 $\rm km~s^{-1}$
instead of 625 $\rm km~s^{-1}$.  Thus for the interesting range of shock speeds, the
neutrals move away from the shock twice as fast as is assumed in the model shown in
Figure \ref{model}.  The relative velocity is much smaller for the second charge transfer
event, so after two charge transfers the neutrals have a speed closer to that of the downstream
plasma.  Kevin Heng (2007, private communication) has provided the average downstream
speed of the broad component neutrals computed by the \citet{hm} model code.  For the
2500 to 3000 $\rm km~s^{-1}$ velocity range it is 1.32 times the plasma speed. Therefore,
for comparison to observations we stretch the spatial scale of the broad component emission
shown in Figure 3 by a factor of 1.32.  Particle conservation implies that the neutral
density is decreased by the same factor, so the broad component emissivity at each point
is reduced by a factor of 1.32.  Finally,
for comparison with the observations we multiply the broad component
emission by 0.75 to account for the drop of sensitivity in the blue wing resulting from
the transmission of the F658N filter.

To model the geometry of the filament, we assume a shock surface that is concave outward
with radius of curvature R.  The H$\alpha$ emissivity from the planar model extends
behind the shock at each point.  Numerical integration then gives the brightness as
a function of position relative to the tangent point of the shock.  For comparison with
the observations, we convolve the model emissivity with the ACS point spread function.
In order to properly account for the pixel sampling in cuts made at about 45$^\circ$
to the rows and columns of the detector, we measured the Gaussian widths of stars near
the filament (2.31 pixel FWHM), then placed a dense series of circular Gaussians along
a 45$^\circ$ line, extracted the profile perpendicular to that line, and measured a
slightly broader 2.43 pixel width.  This Gaussian is convolved with the models for
comparison with the observed spatial profiles.

Figure~\ref{schem} is a schematic diagram of the geometry we imagine in a plane that
includes the line of sight (dashed line) and a radial vector from the center of the
SNR.  The light line shows a large scale ripple in the shock front, and we have superposed
a smaller scale ripple to obtain the shape of the heavier curve that is shaded toward
the inside of SN1006.  The shading indicates the H$\alpha$ emissivity, which fades
gradually from the shock front towards the inside of the SNR.  The trace in the lower
right indicates the H$\alpha$ brightness obtained for a cut across the tangent point
of the shock by integrating the H$\alpha$ emissivity along different lines of sight. 

\subsection{Comparison of Models and Observations}

Figure~\ref{obspred10} compares a grid of models with different $n_0$ and R
to the spatial profile at position 10.  The sharp spikes ahead of the H$\alpha$
peak are faint stars in the extraction region.  The models have been scaled to match the
peak H$\alpha$ brightness of the filament by simply adjusting the neutral fraction.
Neutral fractions above 1 are obviously unphysical, so models with low $n_0$ and
small R are ruled out.  \citet{gha02} estimated a pre-shock neutral fraction of
0.1 from the ratios of He I and He II lines to H$\alpha$, so we take the permitted range
of neutral fractions to be $0.05 \le f_{neut} \le 0.2$.  Figure~\ref{obspred10} shows that models with $n_0$ below
about 0.30 $\rm cm^{-3}$ fall off too slowly behind the shock, while the models with
$n_0$ above about 0.35 $\rm cm^{-3}$ give too sharp a peak.  
The R=$5 \times 10^{16}$ cm model comes closest to 
matching the observed profile.  Models with
smaller R predict H$\alpha$ emission fainter than observed.  We conclude
that $0.3~\le~n_0~\le~0.35$ and $10^{16.5}~\le~R ~<~10^{16.8}$ cm.

Figure~\ref{obspred28} shows the analogous plots for position 28.  The peaks in this
section of the filament are both brighter and broader, with a fairly constant H$\alpha$
intensity ahead of the brightness peak.  None of the models match exactly, probably
because the rippled sheet does not follow the assumed shape of an arc of a circle.
From Figure~\ref{obspred28} we conclude that $n_0$ must be greater than 0.25 $\rm cm^{-3}$
to avoid a long tail toward the inside of the remnant, but that densities above
0.4 produce too sharp a peak.  R must be less than $2 \times 10^{17}$ cm, because 
larger radii of curvature predict a shoulder on the outer side of the spatial profile
that exceeds the observations, while R less than $0.7 \times 10^{17}$ cm requires unacceptably
high values of $f_{neut}$ to match the brightness. We conclude that the acceptable
ranges are $0.25 ~\le~n_0~\le~0.35$ and $0.7 \times 10^{17}~\le~R~\le~1.5 \times 10^{17}$ cm.
We note, however, that the agreement between the model spatial profiles and the
observations is not as good as at position 10.  Based on Figure~\ref{halpha} , it seems possible that
the bright filament at position 28 contains more than one tangency to the line of sight,
broadening the spatial profile and increasing the total brightness.  Thus the upper
limit on $n_0$ and the lower limit on R are less secure than in the case of position 10.

Table 1 presents the ranges of $n_0$ and R derived from other positions along the filament.
We did not attempt to model other positions where the emission peak is faint or where
complex morphology indicates a more complex structure than can be approximated by a simple
curved sheet.  The profiles of positions 8 through 11 are qualitatively similar to that
of position 10, and those of positions 26 through 29 are like that of position 28.

\section{Discussion}
 
\subsection{Pre-shock density}

The densities near 0.3 $\rm cm^{-3}$ that we derive fall in the middle of the range of 
previous values.  \citet{hss86} estimated 0.05-0.1 $\rm cm^{-3}$ from the global X-ray 
spectrum.  Since the pre-shock density in the NE, and quite likely the rest of the remnant, 
is about 2.5 \citep{lon03} to 4 \citep{korreck} times smaller than in the region we observed, the
\citet{hss86} estimate is consistent with our results, even though most of the X-ray emission 
was subsequently shown to be non-thermal in nature.  \citet{win97} obtained a pre-shock
density near 1.0 from the spatial profile of the X-ray emission, but \citet{lon03}
found $n_0~\simeq~0.25$ from the value of $n_e t$ obtained from the Chandra spectrum
of their region NW-1, which lies immediately behind the region of the Balmer line filament
we observed.  The Chandra spectrum showed solar abundances, which is consistent with 
the interpretation of shocked interstellar gas, and the temperature of 0.7 keV is
consistent with very inefficient thermal equilibration between ions and electrons.
Thus the present results agree well with the results of shock wave models of the
X-ray spectra and confirm the parameters derived.  However, as \citet{lon03} point
out, the shock models did not provide an acceptable $\chi$-squared fit to the data,
so some aspect of the physics remains to be understood.  In particular, with $n_0$
as an independently measured quantity rather than a free parameter, reanalysis of
the Chandra spectrum might be able to place better limits on the non-thermal
emission in the NW part of SN1006.

Another density estimate
for the NW region of SN1006 was obtained by \citet{lam96}, who found that the
relative intensities of the UV lines could be explained if only about half the
O VI emission fell within the aperture of the HUT telescope.  This would require
an ionization length for the O VI of about 10$^\prime$$^\prime$, or $n_0~\sim~0.04~\rm cm^{-3}$.
This interpretation is not consistent with the values of $n_0$ derived for the 
same region from the HST image, so we conclude that there is a problem either
with the models of \citet{lam96}, or perhaps more likely with the reddening
correction in the far UV.

A density estimate that is independent of SN1006 itself comes from H I 21 cm
observations by \citet{dub02}.  An H I feature with a column density of
$7 \times 10^{20}~\rm cm^{-2}$ at a velocity of -6 $\rm km~s^{-1}$ lies
just outside the NW rim of SN1006.  While the velocity does not correspond
to the Galactic rotation value at the distance to SN1006 
($-25~\le~V_{LSR}~\le~-16 ~\rm km~s^{-1}$), the SNR is far enough from the 
plane that the velocity could easily differ.  \citet{dub02} estimate a
neutral hydrogen density of 0.5 $\rm cm^{-3}$, which with the low neutral fraction
from \citet{gha02} would imply a total density of 5 $\rm cm^{-3}$, well above
the values we derive.  We conclude that either SN1006 is not interacting with
the cloud identified by Dubner et al., or that the shock has only reached the
low density outskirts of the cloud.

\subsection{Length scales}

The length scale given by the radius of curvature is considerably
shorter than the length of the filament itself.  The more or less
straight portion of the H$\alpha$ filament extends for perhaps 7$^\prime$,
or about 4 pc, roughly 100 times typical radius of curvature derived for the ripples.
The actual line of sight scale of the filament is several
times that of the bright rim at the trailing edge of the
filament.  The region ahead of the trailing bright rim at
position 28 has a nearly constant surface brightness of
$2~-~3 \times 10^{-5}~\rm photons~cm^{-2}~s^{-1}~\arcsec^{-2}$.
For $n_0$ = 0.3, $f_{neut}$ = 0.1 and $V_s$ = 2900 $\rm km~s^{-1}$, 
this implies an angle of 6 to 10 degrees between the shock and 
the line of sight.  The roughly constant intensity region is
about $2.8 \times 10^{17}$ cm in radial exent, so the range of angles
implies a depth along the line of sight between 1.5 and $2.3 \times 10^{18}$ cm,
or about a tenth the scale of the filament in the plane of the sky.
The $10^{18}$ cm scale corresponds to the smoother curve in the schematic
diagram in Figure~\ref{schem}, while the $10^{17}$ cm scale derived from
the H$\alpha$ brightness profiles corresponds to the smaller scale ripple
superposed to obtain the shock geometry shown in the schematic.

Even the larger scale inferred for the direction along the line of sight
is several times smaller than the length of the filament in the plane of 
the sky.  Very faint H$\alpha$ emission,
which is not apparent in Figure~\ref{halpha}, is seen in the very
deep H$\alpha$ image of \citet{wink03} extending out ahead of the
bright filament (their Figure 5).  This is probably a shock in lower density
gas that may be farther from tangency with the line of sight, and it probably 
extends for a distance comparable to the length of the filament.

Figure 2 shows that ripples on the scale of $10^{18}$ cm can be seen in
the direction along the filament, for instance between the two boxes where
radial profiles were extracted.  Ripples on the $10^{17}$ cm scale are
not apparent, perhaps because a $10^{17}$ cm ripple with 10\% amplitude
would be only a few resolution elements radially in Figure 2.  However, such
small scale rippling undoubtedly contributes to the widths of the H$\alpha$
brightness peaks and prevents them from being as sharp as the model peaks,
as is particularly apparent in the position 28 profile.  It is also quite 
likely that the ripples are not isotropic.  If the magnetic field lies
near the plane of the sky in the NE-SW direction, then density structures
would be elongated in that direction and smaller scales would appear along
the line of sight.  There is some evidence that the field does lie in this
direction based on the cap-like morphology of the non-thermal X-ray emission
\citep{wil96}, but there is also a suggestion that the field lies in
the SE-NW direction based on the relative temperatures of protons and oxygen
ions \citep{korreck}. 
 
If we attribute the $\sim~2 \times 10^{18}$ cm scale length of the ripples to
density fluctuations in the interstellar gas, we can estimate the
amplitude of the density fluctuation from the amplitude of the ripple.
The amplitude of the ripple is about one tenth the wavelength,
and that should be about equal to $\delta$V/V.  For constant ram pressure,
$n V^2$, this requires density fluctuations of about 20\%. As was found
for similar ripples in the blast wave of the Cygnus Loop, this agrees reasonably well with
the expectations from the spectrum of interstellar turbulence \citep{minter, ray03}.
However, the smaller scale ripples revealed by the $10^{17}$ cm radius of
curvature also seem to have amplitudes of order 1/10 the wavelength, and
a Kolmogorov spectrum of density fluctuations from the turbulent cascade
would lead one to expect a smaller amplitude \citep{minter}.  \citet{beresnyak}
show that the spectrum of density fluctuations may be considerably flatter than
the Kolmogorov spectrum.

One feature in the filament is probably not due to density fluctuations in the
ISM, however.  The bulge near the western edge of the filament in Figure~\ref{halpha}
is very similar to one farther west along the filament that coincides
closely with a bright knot of X-ray emission with enhanced elemental
abundances \citep{lon03, vink03}.  The X-ray knot (position NW-2 of \citet{lon03})
is clearly a knot of ejecta overtaking the blastwave.  The bulge at the SW
corner of the ACS image is probably a similar structure with lower X-ray contrast
due to its smaller size and perhaps differences in density, ionization time scale
and abundances.  According to hydrodynamic simulations of
Type Ia SNRs, ejecta knots form at the Rayleigh-Taylor unstable contact discontinuity,
but they reach at most 87\% of the blast wave radius \citep{wang01}.  Thus
the ejecta knots should be over a parsec, or about 2$^\prime$ behind the 
Balmer filament.  \citet{warren} have pointed out that if a substantial fraction
of the energy dissipated by a shock goes into accelerating cosmic rays, the 
ejecta knots can come much closer to the outer shock.  In the case of the
bulge in the southwestern corner of the HST image and the X-ray knot farther to the SW,
the lack of synchrotron emission 
at radio and X-ray wavelengths indicates that little energy goes into cosmic rays.
Instead, since the blast wave encounted the dense gas in the NW sector of SN1006 only about
180 years ago \citep{lon03}, the ejecta knots have undoubtedly overtaken the blast wave 
because the blast wave decelerated when it encountered denser gas.
This has an important implication for analyses
of the fraction of shock energy that goes into cosmic rays.  \citet{warren} show
that ejecta knots in Tycho's SNR come much closer to the blast wave than predicted
by \citet{wang01}, and they interpret this in terms of energy that goes into
accelerating particles to high energies.  That argument 
can be applied to most of Tycho's SNR, but 
regions where the SNR shell is flattened or where the Balmer line emission is especially
bright should be avoided in the analysis, because those are regions where the shock has
probably been decelerated by higher density gas.

\subsection{Shock Precursor}

Shock wave precursors have been inferred from anomalously high line widths
of the narrow component H$\alpha$ emission in a number of SNRs and attributed
to heating in the precursor predicted by diffusive shock acceleration models
or to broad component hydrogen atoms that overtake the shock and heat the
upstream gas \citep{smi94, hes94}.  The small width of the narrow
component in SN1006 means that there is no evidence for such a precursor,
but it is nevertheless worthwhile to place on limit on the emission from
such a precursor.  One expects a more or less exponential falloff of brightness
ahead of the shock peak on a scale given by $\kappa / V_s$, where $\kappa$
is the comic ray diffusion coefficient, or by $(n_0 \sigma)^{-1}$ where
$\sigma$ is the charge transfer cross section.  The former would be about
1\arcsec\/  for $\kappa~\sim~10^{25}~\rm cm^2 s^{-1}$. 
For a speed near 3000 $\rm km~s^{-1}$ and the densities in Table 1, the
latter would be about 0.6\arcsec.  The brightness could be a significant
fraction of the narrow component brightness, or up to half the total
brightness of the filament.

The concave outward ripples we have
analyzed so far are not appropriate for searching for a precursor, because
the emission from the curved part of the shock front could easily resemble
emission from a precursor.  Instead the bright rims about 10\arcsec\/ ahead
are convex outward ripples, so that any emission ahead of the peak would be
due to a precursor.  Spatial profiles do show shoulders of order 1\arcsec
wide ahead of the leading bright rims in several places.  However, there is
additional faint emission out ahead of the main filament along the entire
region imaged (\citet{wink03}, Figure 5).  That suggests that the leading 
rim might be {\it S}-shaped along the line of sight rather than a convex 
outward arc, so the $H\alpha$ could be similar to that in the models described 
above (Figures 4 and 5).  In particular, the sections of the filament that show 
an H$\alpha$ shoulder ahead of the peak emission seem to the the regions where 
the diffuse emission ahead of the filament is especially bright.
Therefore, we can only place an upper limit on the precursor just ahead
of the shock of about 1/3 the peak brightness.  Stricter limits of about 1/10 the 
peak brightness can be placed in some sections.  Given the lack of evidence
for heating in a precursor in SN1006, this is not surprising.  It does
suggest, however, that morphology alone may not be enough to establish the
existence of a precursor without some additional information such as [N II]
or [S II] line emission \citep{hes94}, spatial separation of narrow
and broad emission \citep{lee06}, superthermal [N II] line widths \citep{sol03},
or narrow component line widths of order 40 $\rm km~s^{-1}$.

\subsection{Bulk velocity contribution to line widths}

The widths of the H$\alpha$ and UV line profiles have been analyzed under the
assumption that the broadening due to bulk motions in parts of the filament
that are not quite tangent to the line of sight is small.  The range
of angles between the line of sight and the shock surface derived
above for the region between the leading
and trailing rims implies doppler shifts up to 1/6 the post-shock speed,
or $V_s$/8.  Making the extreme assumption that both red- and blue-
shifts are present, and ignoring the contribution of the bright rims
that are tangent to the LOS, the contribution of bulk motions as large 
as 370 $\rm km~s^{-1}$ could be present, or a full width of 750 $\rm km~s^{-1}$.
With measured line widths of 2290 $\rm km~s^{-1}$ for H$\alpha$ \citep{gha02}
and 2100 to 2600 $\rm km~s^{-1}$ for the UV lines \citep{ray95, korreck},
bulk motions contribute at most 6\% to the measured line widths when added
in quadrature to the thermal widths.

\section{Summary}

An HST image of the Balmer line filament in SN1006 resolves the region where neutral
hydrogen is excited and ionized.  The thickness of the ionization zone implies a
total density of 0.25 to 0.40$\rm cm^{-3}$, roughly in the middle of the order of magnitude
range of earlier density estimates.  It agrees well with the density inferred by
\citet{lon03} from the ionization time scale of the X-ray emission just behind the
Balmer line filament, providing confirmation for the shock wave models used to interpret
the X-ray spectra.  Ripples in the filament on parsec scales are consistent with 
those expected for a shock propagating through a medium with $\sim$ 20\% density
fluctuations on a parsec scale, in accordance with measurements of turbulence
in the ISM.  The ripples on $10^{17}$ cm scales have higher amplitude than expected for
a Kolmogorov spectrum of density fluctuations, however. A bulge in the H$\alpha$ 
filament corresponding to a knot of ejecta
shows that ejecta knots can overtake the outer SNR shock when the blast wave 
decelerates upon encountering a dense cloud.

The density estimates in this paper are based upon model calculations for the 
emission from non-radiative
shocks.  The models took recent improvements by \citet{hm} into account in an
approximate way, but a more complete model would be valuable.  Models of the 
3D morphology of the filament that go beyond the simple shape assumed could improve
the accuracy of the derived parameters, especially when spatially resolved spectra become
available to pin down the additional parameters involved. 

\acknowledgments

This work is based on observations made with the NASA HST, and we are grateful
to Max Mutchler of STScI for help in using multidrizzle on the images.  We thank 
Dick McCray and the referee, Kevin Heng, for very helpful
comments, and Kevin Heng for computing the average velocity of the fast
neutrals relative to the shock.  This work is supported
by HST Grant GO-10577.01-A to the Smithsonian Observatory. 
This work made use of the NASA Astrophysics Data System (ADS).

{\it Facilities:} \facility{HST}

\clearpage


\begin{deluxetable}{rllccc}
\tablecaption{Densities and radii of curvature from H$\alpha$ spatial profiles}
\tablehead{
\colhead{Position} & \colhead{RA$_{2000}$} & \colhead{$\delta_{2000}$} & PA & \colhead{$n_0$} & \colhead{R$^a$}
} 
\startdata
    8 &  $15^h 02^m 23.488^s$ & $-41^\circ 44\arcmin 12.45\arcsec$ & 152.15 &  $>$0.35     &  $<$0.4     \\
    9 &  $15^h 02^m 23.136^s$ & $-41^\circ 44\arcmin 14.62\arcsec$ & 154.65 &  $>$0.40     &  $<$0.4     \\
   10 &  $15^h 02^m 22.796^s$ & $-41^\circ 44\arcmin 16.96\arcsec$ & 154.65 & 0.30 - 0.35  & 0.3 - 0.6 \\
   11 &  $15^h 02^m 22.450^s$ & $-41^\circ 44\arcmin 19.29\arcsec$ & 155.90 &  $>$0.35    &  $<$0.5    \\
   26 &  $15^h 02^m 17.340^s$ & $-41^\circ 44\arcmin 55.13\arcsec$ & 153.40 &  0.25 - 0.35 &  2 - 5     \\
   27 &  $15^h 02^m 17.014^s$ & $-41^\circ 44\arcmin 57.91\arcsec$ & 154.65 & 0.30 - 0.40 &  0.8 -  1.2\\
   28 &  $15^h 02^m 16.682^s$ & $-41^\circ 45\arcmin 00.34\arcsec$ & 153.40 & 0.25 - 0.35  & 0.7 - 1.5  \\
   29 &  $15^h 02^m 16.340^s$ & $-41^\circ 45\arcmin 02.74\arcsec$ & 157.15 & 0.35 - 0.40 &  0.5 - 1   \\

\enddata
\tablenotetext{a}{$10^{17}$ cm}
\end{deluxetable}

\clearpage



\begin{figure}
\plotone{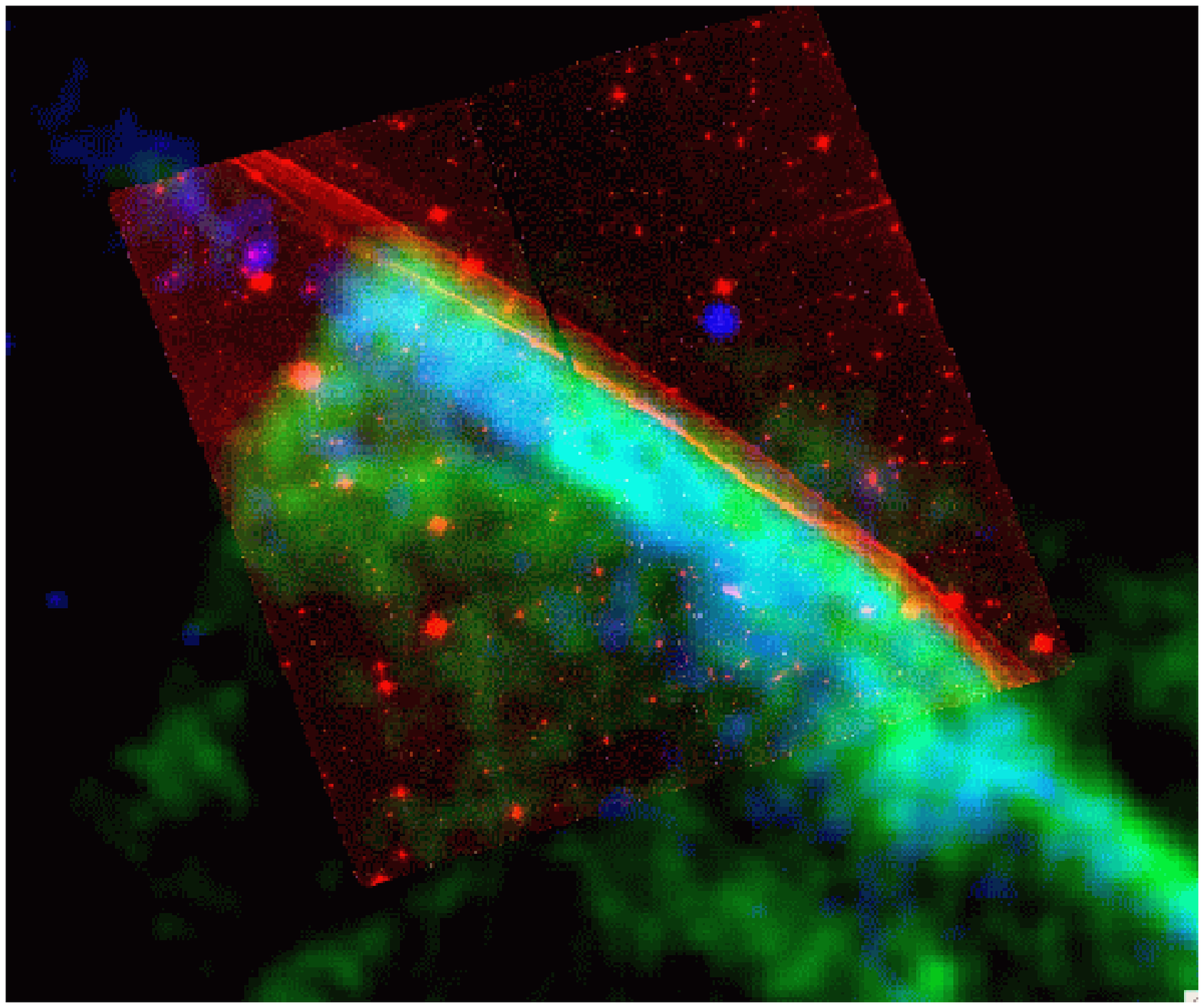}
\end{figure}

\begin{figure}
\plotone{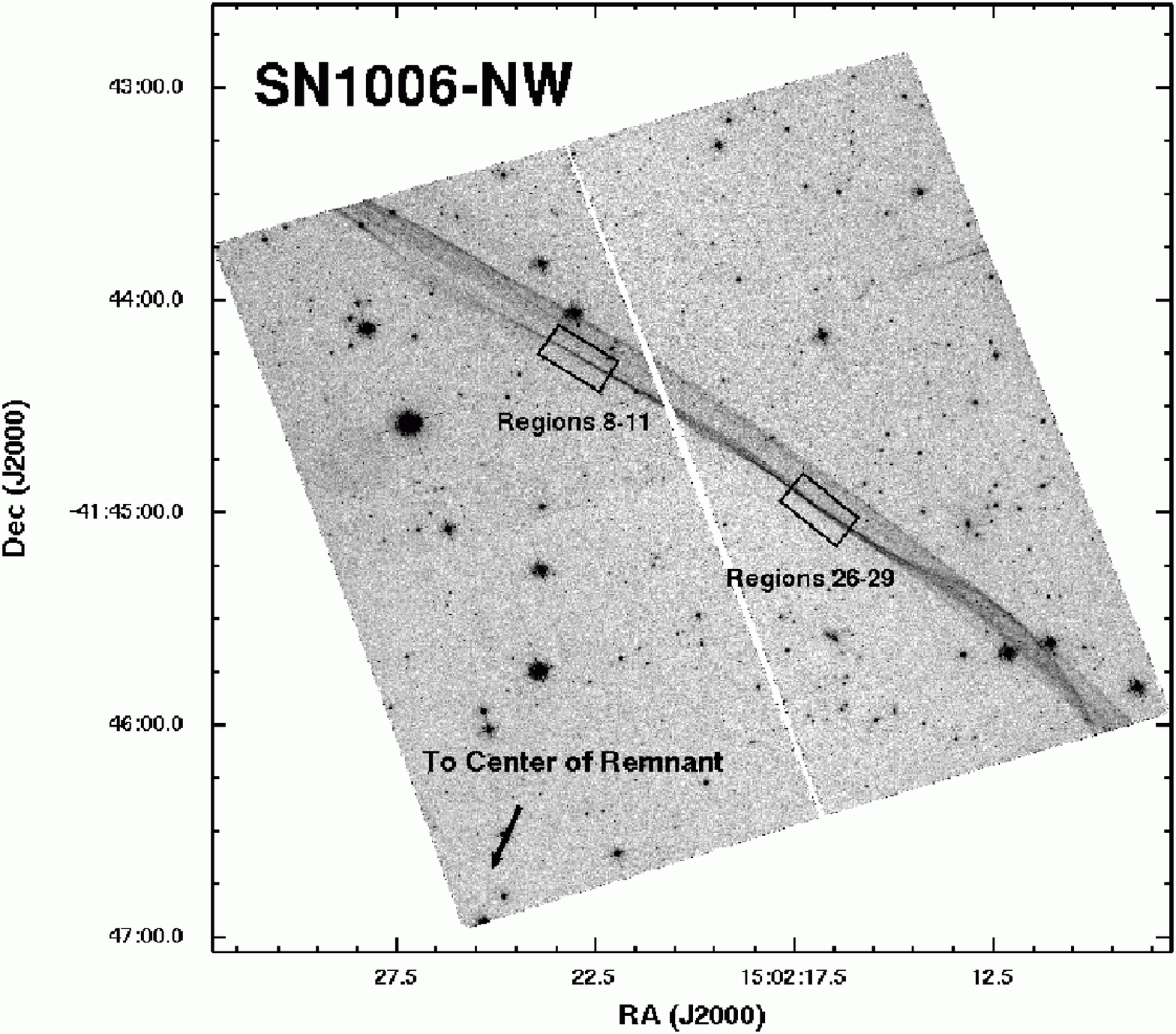}
\end{figure}

\begin{figure}
\plotone{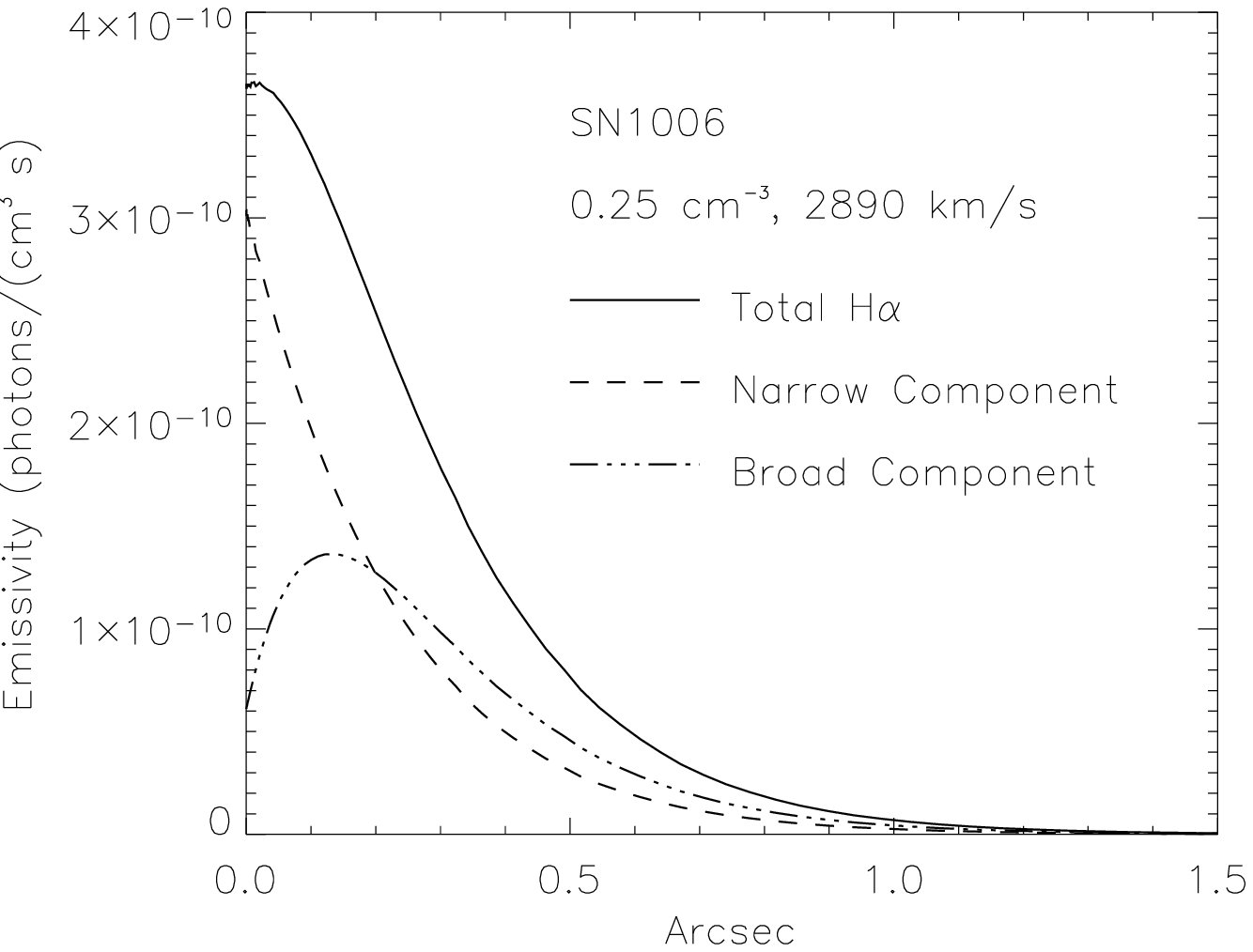}
\end{figure}

\begin{figure}
\plotone{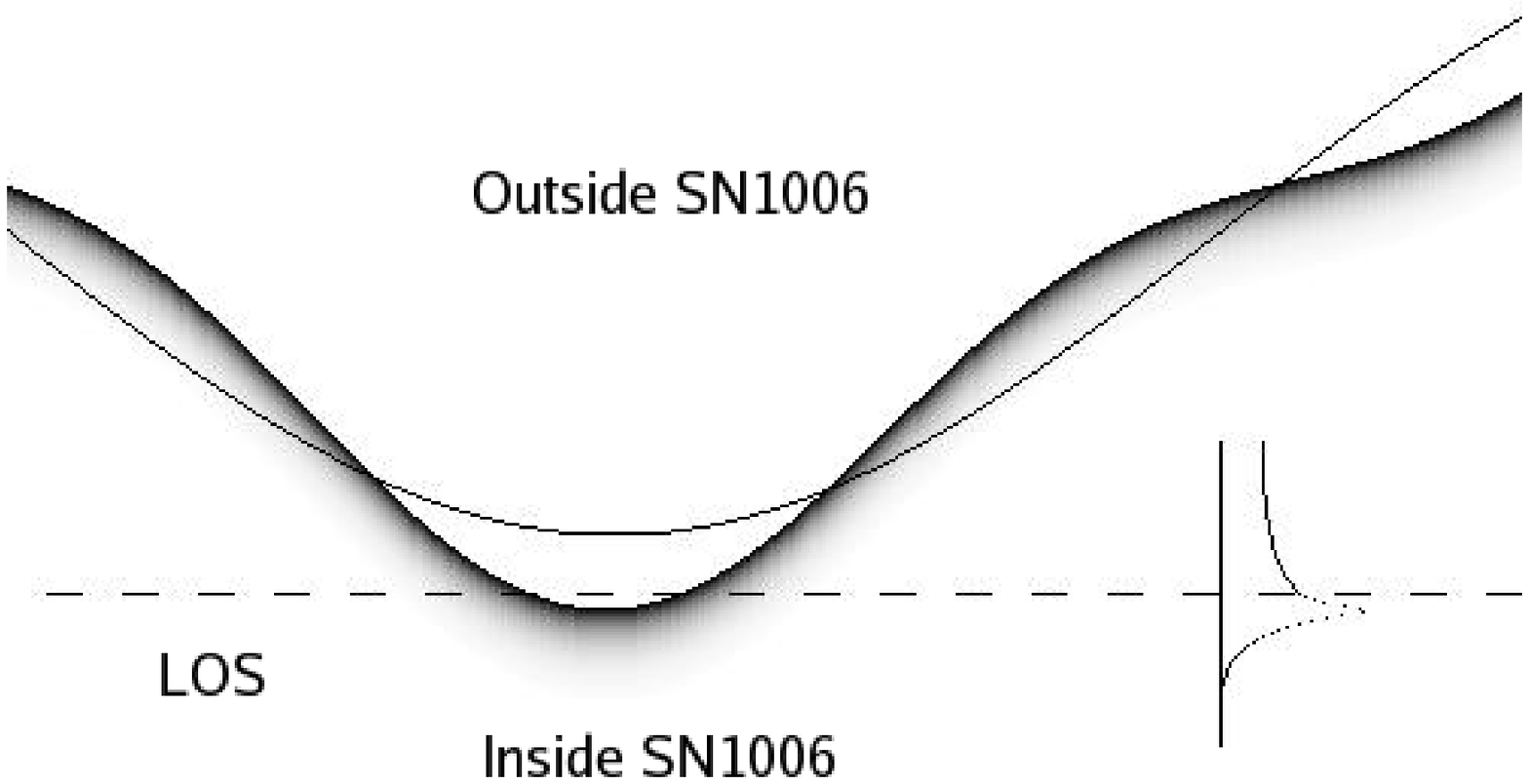}
\end{figure}

\begin{figure}
\plotone{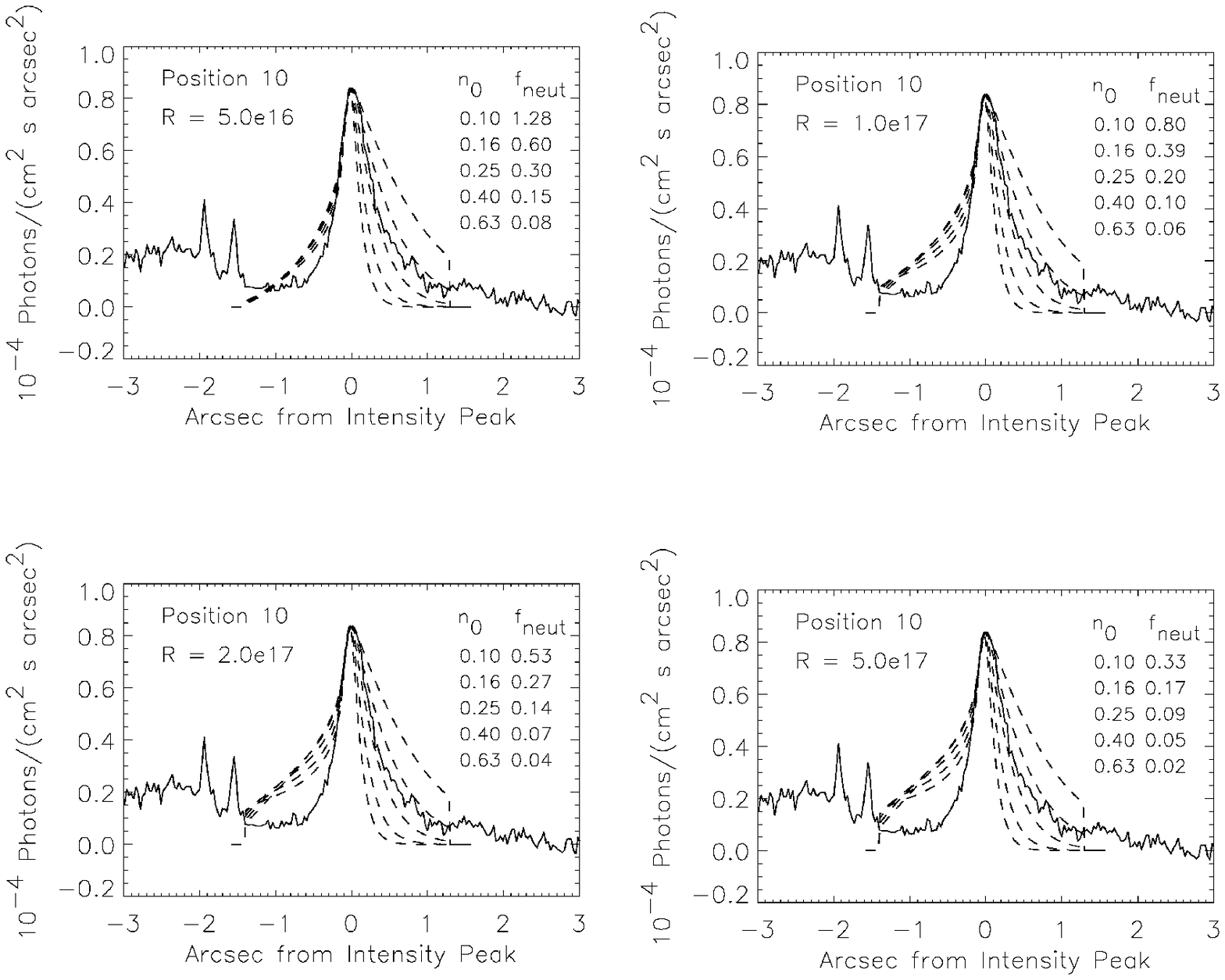}
\end{figure}

\begin{figure}
\plotone{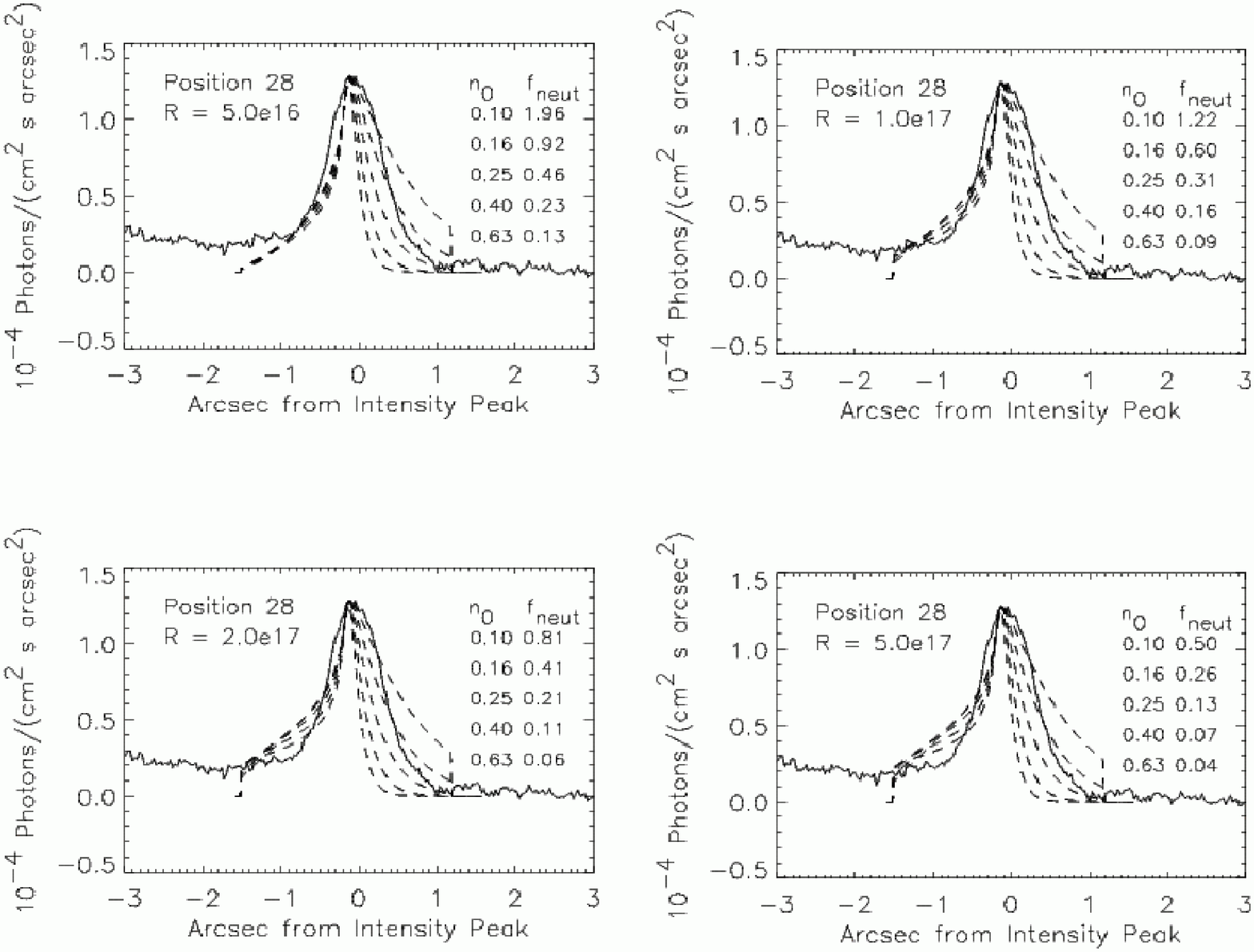}
\end{figure}

\clearpage

\figcaption[f1.ps]{A three-color figure showing HST/ACS F658N (in red) and two energy bands
of Chandra data (0.4 - 0.7 keV in green and 0.8 - 1.5 keV in blue).  The field of view
shown is 279\arcsec\ $\times$ 331\arcsec, with north up and east to the left.  The X-ray
data have been binned by 4 pixels and smoothed with a 3-pixel Gaussian filter to remove
pixelation.  Note the extended region of X-rays coincident with the bulge in the H$\alpha$
image.  The drop-off in X-ray at upper left is due to a CCD-chip boundary and is not real. 
\label{xrayhalpha}}

\figcaption[f2.ps]{The full field of view ACS F658N (H$\alpha$) image
of the NW Balmer filament in SN 1006.  Two 10\arcsec\ $\times$
20\arcsec\ boxes indicate the filament tangencies discussed in detail
in this paper. 
\label{halpha}}

\figcaption[f3.eps]{H$\alpha$ emissivity as a function of position behind the shock for
a non-radiative shock with $n_0$ = 0.25 $\rm cm^{-3}$, $f_{neut}$ = 0.1 and
$T_e /T_p$ = 0.05 at the shock.  This model was computed under the assumption that charge
transfer is isotropic.  To account for the effects described by \citet{hm} we stretched
out the broad component emission by a factor of 1.5 when comparing with the observations.
The scale assumes that 1\arcsec = $3.1\times 10^{16}$ cm for a distance of 2.1 kpc.
\label{model}}

\figcaption[f4.ps]{Schematic diagram of the geometry of the emitting filament along the
line of sight (LOS).  The light line shows a large scale ripple in the shock front,
and the darker curve is the result of adding a smaller scale ripple to the light line.
The fading of the dark curve towards the inside of SN1006 indicates the falloff
of emissivity with distance behind the shock (Figure~\ref{model}).  The trace in
the lower right shows the H$\alpha$ brightness as a function of position for lines
of sight passing near the tangency to the LOS.  It is obtained by integrating
the H$\alpha$ emissivity along each line of sight.
\label{schem}}

\figcaption[f5.ps]{Comparison of observed and computed H$\alpha$ spatial profiles for position 10.
The shock wave is concave outward, with the assumed radius of curvature shown in each panel.
The inside of the remnant is to the right.
The dashed curves correspond to the pre-shock densities listed, with the lowest density giving
the highest curve on the inside and the lowest on the outside.  All the models have been scaled
to the observed peak intensity by using the neutral fraction shown.  Neutral fractions above
1 are obviously unphysical, implying that that combination of pre-shock density and radius
of curvature cannot account for the observed brightness.
\label{obspred10}}

\figcaption[f6.ps]{Same as Figure 4 for position 28.
\label{obspred28}}

\end{document}